\documentclass[aps,prc,twocolumn,superscriptaddress,showpacs]{revtex4}
\usepackage{graphicx}
\usepackage{sidecap}

\newcommand{\aNSCL}{\affiliation{National Superconducting Cyclotron Laboratory, Michigan
State University, East Lansing, MI, USA 48824}}
\newcommand{\aDubna}{\affiliation{Flerov Laboratory of Nuclear Reactions, JINR,
141980 Dubna, Moscow Region, Russian Federation}}
\newcommand{\aMSUphys}{\affiliation{Department of Physics and Astronomy, Michigan State
University, East Lansing, MI, USA 48824}}
\newcommand{\aMSUchem}{ \affiliation{Department of Chemistry, Michigan State University,
East Lansing, MI, USA 48824}}
\newcommand{\aRIKEN}{\affiliation{RIKEN Nishina Center,
RIKEN, Wako-shi, Saitama 351-0198, Japan}}

\newcommand{\softsh}[1]{\texttt{#1}}
\newcommand{\soft}[1]{\texttt{#1 }}
\newcommand{\lise}{\soft{LISE}}
\newcommand{\lisepp}{\soft{LISE$^{++}$}}
\newcommand{\liseppsh}{\softsh{LISE$^{++}$}}

\newcommand{\epax}{\soft{EPAX}}
\newcommand{\epaxsh}{\softsh{EPAX}}

\begin{document}


\title{Production of very neutron-rich nuclei with a $^{76}$Ge beam}


\author{O.~B.~Tarasov}\email[]{tarasov@nscl.msu.edu} \aNSCL \aDubna
\author{M.~Portillo} \aNSCL
\author{A.~M.~Amthor} \aNSCL \aMSUphys
\author{T.~Baumann} \aNSCL
\author{D.~Bazin} \aNSCL
\author{A.~Gade} \aNSCL \aMSUphys
\author{T.~N.~Ginter} \aNSCL
\author{M.~Hausmann}\aNSCL
\author{N.~Inabe} \aRIKEN
\author{T.~Kubo} \aRIKEN
\author{D.~J.~Morrissey} \aNSCL \aMSUchem
\author{A.~Nettleton}   \aNSCL \aMSUphys
\author{J.~Pereira} \aNSCL
\author{B.~M.~Sherrill}\aNSCL \aMSUphys
\author{A.~Stolz} \aNSCL
\author{M.~Thoennessen}\aNSCL \aMSUphys

\date{\today}

\begin{abstract}

Production cross sections for neutron-rich nuclei from the
fragmentation of a $^{76}$Ge beam at 132~MeV/u were measured. The
longitudinal momentum distributions of 34 neutron-rich isotopes of
elements  \protect{$13\le Z\le 27$} were scanned using a novel
experimental approach of varying the target thickness. Production
cross sections with beryllium and tungsten targets were determined
for a large number of nuclei including 15 isotopes first observed in
this work. These are the most neutron-rich nuclides of the elements
\protect{$17\le Z\le 25$} ($^{50}$Cl, $^{53}$Ar, $^{55,56}$K,
$^{57,58}$Ca, $^{59,60,61}$Sc, $^{62,63}$Ti, $^{65,66}$V, $^{68}$Cr,
$^{70}$Mn). A one-body $Q_{\text{g}}$ systematics is used to
describe the production cross sections based on thermal evaporation
from excited prefragments. Some of the fragments near $^{58}$Ca show
anomalously large production cross sections.
\end{abstract}

\pacs{25.70.Mn, 27.40.+z, 27.50.+e}

\maketitle



\section{Introduction\label{Intro}}

The discovery of new nuclei in the proximity of the neutron dripline
provides a benchmark for nuclear mass models, and hence for the
understanding of the nuclear force and the creation of elements.
Once the production methods and cross sections for the formation of
neutron-rich nuclei are understood, further investigations to study
the nuclei themselves, such as decay spectroscopy, may be planned.
Therefore, the determination of production rates for the most exotic
nuclei continues to be an important part of the experimental program
at existing and future rare-isotope facilities.

A number of production mechanisms have been used to produce
neutron-rich isotopes for  \protect{$17\le Z\le 25$}.  For example,
$^{57,58}$V were observed for the first time in deep-inelastic
reactions in 1980~\cite{BRE-PRC80}, $^{59,60}$V~\cite{DGM-ZPA85} and
$^{61}$V~\cite{WEB-ZPA92} by fragmentation of a $^{86}$Kr beam for
the first time in 1985 and 1992, respectively, and $^{62,63,64}$V
were first observed  as products of projectile fission in
1997~\cite{MB-PLB97}. The on-line isotope separation  technique was
used to produce for the first time $^{53,54}$K and $^{53}$Ca in
1983~\cite{LAN_PLB83}, and 14 years later, $^{54,55,56}$Ca isotopes
were observed following in-flight fission of
$^{238}$U~\cite{MB-PLB97}. The possibility of producing very
neutron-rich isotopes (e.g. $^{58,60}$Ca) in transfer-type reactions
at intermediate beam energies has been discussed
recently~\cite{AG-PRC08}.

New progress in the production of neutron-rich isotopes is possible
given the increased primary beam intensities at the National
Superconducting Cyclotron Laboratory (NSCL) at Michigan State
University and advances in experimental techniques \cite{TB-N07}
allowed probing further into unknown regions of the table of the
isotopes with fragmentation reactions. Indeed, recent measurements
at the NSCL~\cite{OT-PRC07,TB-N07,PFM-BAPS08} have demonstrated that
the fragmentation of $^{48}$Ca and $^{76}$Ge beams can be used to
produce new isotopes in the proximity of the neutron dripline.
Continuing this work, we report here the next step towards the
fundamental goal of defining the absolute mass limit for chemical
elements in the region of calcium.

\begin{figure} [t]
\includegraphics[width=1.0\columnwidth]{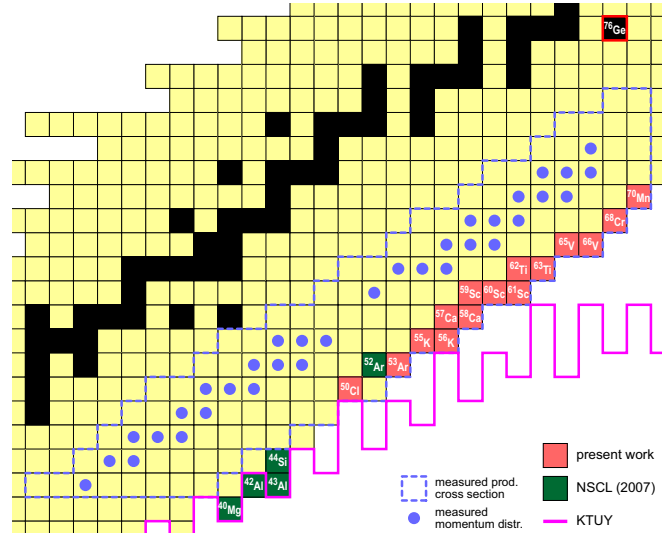}
\caption{(Color online) The region of the nuclear chart investigated
in the
  present work. The solid line shows the limit of bound nuclei from the
KTUY mass model~\cite{KTUY-PTP05}. The 15 new isotopes observed for
the first time in the present work are marked by red
squares.\label{chart}}
\end{figure}

The search for new neutron-rich isotopes was carried out with a
primary beam of $^{76}$Ge and using the
 two-stage fragment separator technique that
enabled the discovery of  $^{40}$Mg and $^{42}$Al in
2007~\cite{TB-N07}. In the present measurement, 15 neutron-rich
isotopes with \protect{$33\le N\le 45$} were identified for the
first time (see Fig.\ref{chart}) and indications for the existence
of a new island of inversion centered on $^{62}$Ti --- as predicted
by Brown~\cite{BAB-PPNP01} --- were described in a previous brief
presentation of this work~\cite{OT-PRL09}. In this paper, we
describe the details of our  experimental approach and discuss the
results.


\begin{SCfigure*}
\includegraphics[width=0.6\textwidth]{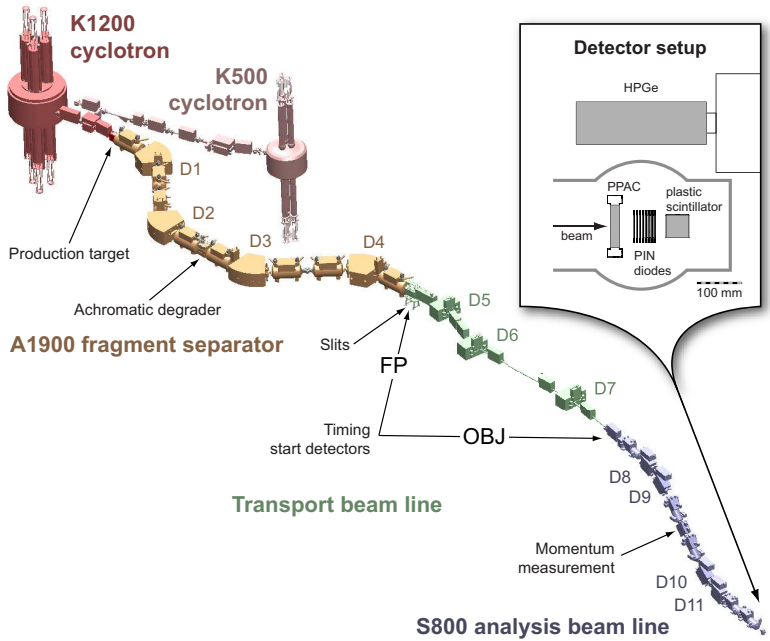}
\caption{(Color online) Sketch of the experimental setup at the
NSCL. This overview   diagram shows the equipment used for the
production,
separation and   detection of new neutron-rich nuclei.\\  \\
 \label{Fig_setup}}
\end{SCfigure*}

\section{Experimental Details\label{secExpt}}
\subsection{Setup\label{secSetup}}

A 132 MeV/u $^{76}$Ge beam, accelerated by the coupled cyclotrons at
the NSCL, was used to irradiate a series of beryllium targets and a
tungsten target, each placed at the target position of the A1900
fragment separator~\cite{DJM-NIMA03}. Previous
studies~\cite{OT-PRC07} demonstrated that using the A1900 fragment
separator as the sole separation stage significantly limits the
sensitivity of the experimental setup for very rare isotopes due to
\begin{itemize}
  \item  excessive rates and large pile-up probability in the detector located at the
intermediate dispersive focal plane of the A1900 fragment separator
(required to measure the momentum),
  \item high rates of light nuclei in the
final focal plane detectors that cannot be rejected by the
energy-loss technique.
\end{itemize}

 The combination of the A1900 fragment separator with
the S800 analysis beam line~\cite{DB-NIMB03} to form a two-stage
separator system as described in Ref.~\cite{TB-N07}, allows a high
degree of rejection of unwanted reaction products  so that the full
primary beam intensity can be utilized, while preserving the
unambiguous identification of single exotic ions. The first stage of
the system serves as a selector whereas the second stage contains
detectors and functions as  analyzer, see Fig.~\ref{Fig_setup}.

In the present work the momentum acceptance of the A1900 was varied
from $\Delta p/p = 0.1\%$ to 5\%. For all settings, the horizontal
and vertical angular acceptances were $\pm 65$ and $\pm 40$ mrad. At
the end of the S800 analysis beam line, the particles of interest
were stopped in a telescope of eight silicon PIN diodes
(50$\times$50~mm$^2$) with a total thickness of 8.0~mm. This
provided multiple energy-loss measurements and thus a redundant
determination of the nuclear charge of each fragment in addition to
the total kinetic energy. A 5~cm thick plastic scintillator
positioned behind the Si-telescope served as a veto detector against
reactions in the Si-telescope and provided a measurement of the
residual energy of lighter ions that were not stopped in the
Si-telescope. A position sensitive parallel plate avalanche counter
(PPAC) was located in front of the Si-telescope. This position
measurement allowed for an additional discrimination against various
scattered particles.

The primary $^{76}$Ge$^{30+}$ beam current was continuously
monitored by a
 BaF$_2$ crystal mounted on a photomultiplier tube near the
target position, as well as a non-intercepting capacitive pick-up
before the target. These indirect monitors were  normalized to
Faraday cup readings throughout the course of the experiment. The
average beam intensity for the measurements of the most exotic
fragments was approximately 1~e$\mu$A.

\begin{table}[h]
\caption{Flight paths between various timing
detectors.}\label{Tab_tof}
\begin{tabular}{|l|l|l|c|}
\hline   ~N~ & \multicolumn{ 2}{|c|}{~Timing signals between} & Length (m) \\
\hline   ~1 & ~FP scintillator & 2$^{nd}$ ~PIN detector &      46.0 \\
\hline   ~2 & ~OBJ scintillator & 3$^{rd}$ ~PIN detector &      21.0 \\
\hline   ~3 & ~FP scintillator & ~OBJ scintillator &      25.1 \\
\hline   ~4 & ~Target$^*$ & 3$^{rd}$ ~PIN detector &      81.5 \\
\hline \multicolumn{ 4}{l} {\it{\footnotesize{$^*$ - from the
arrival time relative }}}     \\
\multicolumn{ 4}{l} {\it{\footnotesize{to the phase of the cyclotron rf-signal}}}     \\

\end{tabular}
\end{table}

The time of flight (TOF) of each particle that reached the detector
stack was measured in four different ways (see Table~\ref{Tab_tof}).
All TOF signals were used for the rejection of unwanted events. The
first three TOF measurements listed in the table were used to deduce
the mass-to-charge ratio $A/q$ with a better resolution than with a
single TOF measurement.

The magnetic rigidity ($B\rho$) of each particle was obtained by
combining position measurements from two PPACs located in the
dispersive plane of the S800 analysis beam line with NMR
measurements of the dipole fields.


\begin{table*}[t]
\caption{ Experimental settings}\label{Tab_runs}
\begin{tabular}{|c|c|ccccc|rl|rl|c|c|c|c|c|c|}
\hline
       Data & {\small{Fragment}}  &     \multicolumn{ 5}{|c|}{Magnetic rigidity, $B\rho (Tm)$} & \multicolumn{ 2}{|c|}{Target} &  \multicolumn{
       2}{|c|}{Stripper}
& Wedge  &   $\Delta p/p$ & Time &  Beam  &  Goal \\

           set &  {\footnotesize{of interest}} &     $D_1D_2$ &     $D_3D_4$ &      $D_5D_6D_7$ &     $D_8D_9$ &      $D_{10}D_{11}$ &
            \multicolumn{ 2}{|c|}{\footnotesize{$mg/cm^2$}} &  \multicolumn{ 2}{|c|} {\footnotesize{$mg/cm^2$}} &     {\footnotesize{$mg/cm^2$}} &       (\%) &
                  {\footnotesize{$hour$}} &         particles &            \\
\hline

         1 &        $^{43}$S &     ~4.3233 &     ~4.3233 &     ~4.3134 &     ~4.3036  &       ~4.3000 &         ~Be &        9.8 &    &     -   &      -      &        0.1 &       0.98 &        {\footnotesize{1.41e14}} & \multicolumn{ 1}{|c|}{} \\

         2 &        &      &      &      &     &        &         Be &       97.5 &     &   -    &     -       &        0.1 &       2.03 &         {\footnotesize{1.50e14}} & \multicolumn{ 1}{|c|}{\footnotesize{momentum}} \\

         3 &        &      &      &      &     &        &         Be &        191 &     &    -   &      -      &        0.1 &       1.98 &         {\footnotesize{2.07e14}} & \multicolumn{ 1}{|c|}{\footnotesize{distribution}} \\

         4 &        &      &      &      &     &        &         Be &        288 &      &    -  &      -      &        0.1 &       1.85 &       {\footnotesize{1.17e14}} & \multicolumn{ 1}{|c|}{\footnotesize{}} \\

         5 &        &      &      &      &     &        &         Be &        404 &       &   -  &      -      &        0.1 &       0.94 &         {\footnotesize{5.76e13}} & \multicolumn{ 1}{|c|}{} \\
\hline
         6 &        $^{43}$S &     4.3372 &     4.3233 &     4.3134 &     4.3036 &        4.3000 &         Be &        191 &        &  -  &        20.2 &          1 &       3.36 &     {\footnotesize{1.25e15}} & \multicolumn{ 1}{|c|}{$^{43m}$S} \\

         7 &        &        &     &      &     &       &         Be &        288 &        &  -  &        20.2 &          2 &       5.97 &       {\footnotesize{1.96e15}} & \multicolumn{ 1}{|c|}{\footnotesize{production}} \\
\hline
         8 &        $^{66}$V &     4.3609 &     4.3383 &     4.3222 &     4.3059 &        4.3000 &         Be &        404 &       &  -   &        20.2 &          5 &      19.86 &        {\footnotesize{1.42e16}} & \multicolumn{ 1}{|c|}{} \\

         9 &         &       &      &      &      &     &                 W &        567 &     C & 12 &        20.2 &          5 &      20.74 &        {\footnotesize{1.47e16}} & \multicolumn{ 1}{|c|}{\footnotesize{production}} \\

        10 &       $^{54}$Ar &     4.3529 &     4.3332 &     4.3192 &     4.3051 &        4.3000 &          W &        567 &   Be & 145 &        20.2 &          5 &      10.33 &      {\footnotesize{7.55e15}} & \multicolumn{ 1}{|c|}{\footnotesize{of new }} \\

        11 &         &     &      &      &      &     &                 Be &        629 &       &  -   &        20.2 &          5 &       9.37 &        {\footnotesize{6.15e15}} & \multicolumn{ 1}{|c|}{\footnotesize{isotopes}} \\

        12 &       $^{59}$Ca &     4.3566 &     4.3355 &     4.3205 &     4.3054 &        4.3000 &         Be &        629 &       &  -   &        20.2 &          5 &      21.60 &        {\footnotesize{1.58e16}} & \multicolumn{ 1}{|c|}{} \\

        13 &       $^{58}$Ca &     4.3546 &     4.3342 &     4.3198 &     4.3052 &        4.3000 &          W &        567 &  Be & 97.5 &        20.2 &          5 &      46.59 &     {\footnotesize{3.34e16}} & \multicolumn{ 1}{|c|}{} \\
\hline
\end{tabular}
\end{table*}

The simultaneous measurement of multiple $\Delta E$ signals, the
magnetic rigidity, the total energy, and the TOFs of each particle
provided an unambiguous identification of the atomic number, charge
state and mass number of each ion.  The detection system and
particle identification was calibrated with the primary beam and
confirmed by the locations of gaps corresponding to unbound nuclei
in the particle identification spectrum. In addition, a germanium
$\gamma$-ray detector  was placed near the Si-telescope (see
Fig.\ref{Fig_setup}) to provide an independent verification of the
isotope identification via isomer tagging as described in
Ref.\cite{RG-PLB95}.

\subsection{Experimental runs\label{secPlanning}}

The present experiment consisted of three  parts that are summarized
in Table \ref{Tab_runs}. During all runs, the magnetic rigidity of
the last two dipoles was kept constant at 4.3~Tm while the
production target thickness was changed to map the fragment momentum
distributions. This approach  greatly simplified the particle
identification during the scans of the parallel momentum
distributions.

The   momentum acceptance of the A1900 fragment separator was
restricted to $\Delta  p/p = 0.1\%$ for the first part of the
experiment devoted to the measurement of differential momentum
distributions. The use of different beryllium target thicknesses
(9.8, 97.5, 191, 288, 404~mg/cm$^2$) allowed to cover different
spans of the fragment momentum distributions necessary to extract
production cross sections and also resulted in more isotopes in the
particle identification spectrum. The particle selection depends
critically on the magnetic rigidity of the fragment separator
because the average energy (and therefore the magnetic rigidity) of
a given reaction product decreases  with increasing target
thickness. As a result, the isotopic selection of this system moved
toward more neutron-rich fragments with increasing target thickness.

 For the second part of the experiment, a Kapton wedge
with a thickness of 20.2 mg/cm$^2$ (indicated  as achromatic
degrader in Fig.\ref{Fig_setup}) was used at the dispersive image of
the A1900 to reject less exotic fragments with  an 8~mm aperture in
the focal plane while the separator was set for $^{43}$S. The goal
of this setting was to confirm the particle identification by isomer
tagging with $^{43m}$S~($E_{\gamma}=319$~keV, $T_{1/2}=0.48$~$\mu$s)
and $^{32m}$Al~($E_{\gamma}=735$~keV, $T_{1/2}=0.2$~$\mu$s).

In the final part of the experiment, dedicated to the  search  for
the new isotopes, four settings were  used to cover the most
neutron-rich isotopes  with \protect{$13\le Z\le 27$}, as it is
impossible to find a single target thickness and magnetic rigidity
to produce all of the fragments of interest. Each setting was
characterized by a fragment on which the separator was tuned. A
search for the most exotic nuclei in each setting was carried out
with Be and W targets. The four settings were centered on $^{54}$Ar,
$^{58,59}$Ca and $^{66}$V respectively, and were based on
\liseppsh~\cite{OT-NIMB08} calculations using the parameterizations
of the momentum distributions obtained in the first part of
experiment (see Section~\ref{secMomentum}). The momentum acceptance
of the A1900 was set to the maximum of $\Delta p/p = 5.0\%$ for
these production runs. In contrast to previous studies with the
A1900 as single-stage separator, e.g.~\cite{OT-PRC07},  the present
overall experimental efficiency was dominated by the acceptance of
the two-stage separator and not by the limitations due to high
counting rates of unwanted nuclei.

\section{Analysis of experimental data \label{secAnalysis}}

\begin{figure}
\centering
\includegraphics[width=0.8\columnwidth]{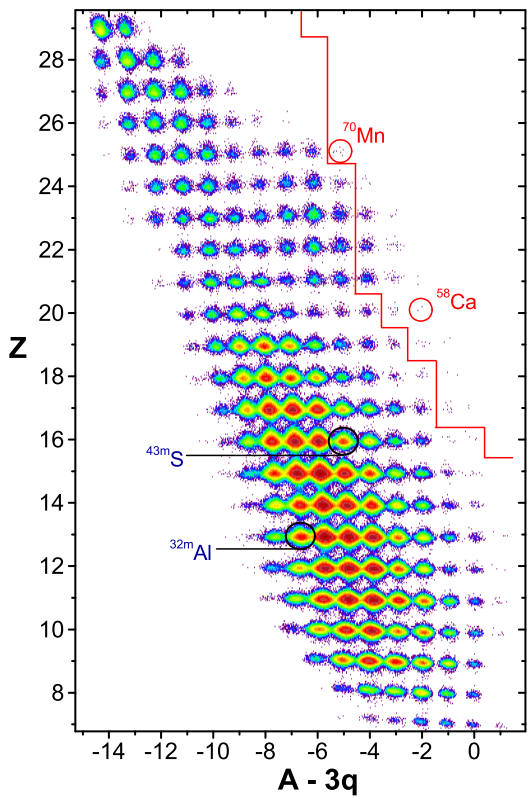}%
\caption{(Color online) Particle identification plot showing the
measured atomic number, $Z$, versus the calculated function $A-3q$
for the nuclei observed in this work ($q$ is the ionic charge of the
fragment). See text for details. The limit of previously observed
nuclei is shown by the solid line as well as the locations of
several reference nuclei. Particle identification was confirmed by
registering gamma-rays from short-lived isomeric states of $^{43}$S
and $^{32}$Al (marked by black circles).\label{Fig_pid}}
\end{figure}

\begin{figure}
\includegraphics[width=0.95\columnwidth]{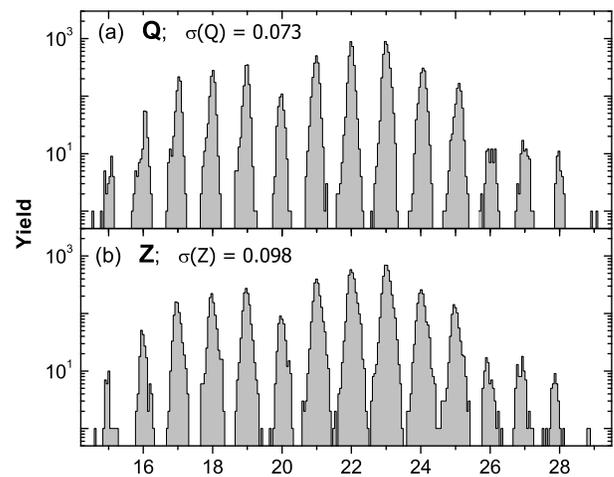}%
\caption{ Ionic (top) and elemental (bottom) spectra obtained with
Eq.~\ref{Eq_Q} and \ref{Eq_Z} for all particles stopped in the
Si-telescope during the data set dedicated to the search for
new isotopes (see Table~\ref{Tab_runs}). \label{Fig_ZQ}}
\end{figure}

\begin{SCfigure*}
\includegraphics[width=0.79\textwidth]{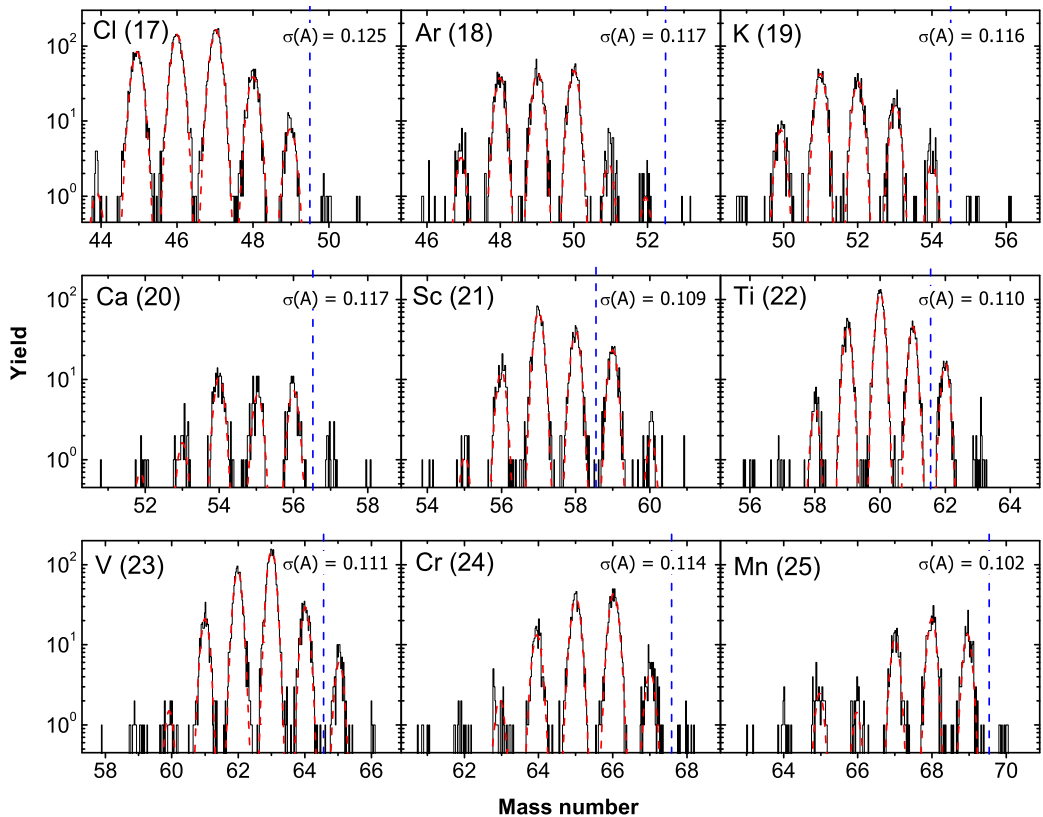}%
\caption{(Color online) Mass spectra of the elements \protect{$17\le
Z\le 25$}. All particles that were stopped in the Si-telescope
during the production runs were analyzed. The limits of previously
observed nuclei are shown by the vertical dashed lines.

Standard deviations produced in multi-peak fits with Gaussian
distributions at constant width (red dashed curves) are shown in the
plots for each element.
\\
\\
\label{Fig_Mass}}
\end{SCfigure*}


The advantage of our approach --- keeping the magnetic rigidity
($B\rho_{0f}$) constant while varying the target thickness
--- can be seen in Fig.~\ref{Fig_pid}, which shows the total
distribution of fully-stripped reaction products observed in all
runs of this work. The range of fragments is shown as the measured
atomic number, $Z$, plotted versus the calculated quantity $A-3q$.
The identification of the individual isotopes in Fig.~\ref{Fig_pid}
was confirmed via isomer tagging using the known isomeric decays in
$^{32}$Al and $^{43}$S, and from holes in the $A-3Z=1$ line
corresponding to the unbound nuclei $^{25}$O and $^{28}$F located
between the particle bound nuclei $^{22}$N and $^{31}$Ne. The
details of the calculation of the particle identification are given
in the Appendix.

The ionic ($q$) and elemental ($Z$) spectra obtained for particles
that stopped in the Si-telescope are shown in Fig.~\ref{Fig_ZQ}. The
peaks were fitted with Gaussian distributions of constant width. The
standard deviations were found to be $\sigma_q = 0.073$ for the
ionic spectrum and $\sigma_Z = 0.098$ for the elemental spectrum,
respectively. This means that the probabilities of one event being
misidentified as a neighboring charge state or element are equal to
$P_q(0.5) = 3.7~10^{-12}$ and $P_Z(0.5)=1.7~10^{-7}$, respectively.

The mass spectra for the isotopic chains from chlorine to manganese
measured during the production runs are shown in
Fig.~\ref{Fig_Mass}. Only nuclei that stopped in the Si detector
stack are included in this analysis. The observed fragments include
15 new isotopes that are the most neutron-rich nuclides yet observed
of elements \protect{$17\le Z\le 25$} ($^{50}$Cl, $^{53}$Ar,
$^{55,56}$K, $^{57,58}$Ca, $^{59,60,61}$Sc, $^{62,63}$Ti,
$^{65,66}$V, $^{68}$Cr, $^{70}$Mn). The new neutron-rich nuclei
observed in this work are those events to the right of the solid
line in Fig.~\ref{Fig_pid} and to the right of the vertical dashed
lines in Fig.~\ref{Fig_Mass}. As noted before~\cite{OT-PRL09}, the
previously reported observation of $^{51}$Cl~\cite{ML-ZPA90} may
have actually been an observation of the hydrogen-like charge state
$^{48}$Cl$^{16+}$.

\section{Results and Discussion\label{secRes}}

\subsection{Parallel momentum distributions\label{secMomentum}}

It is important to be able to predict the momentum distributions of
residues when searching for new isotopes in order to optimize the
fragment separator at the maximum production rate. Also, the
accurate prediction of the momentum distributions allows estimation
of the transmission and efficient rejection of strong contaminants.
Several studies  of the parallel momentum distributions have been
made, e.g.~\cite{AG-PLB74,DJM-PRC89,RP-PRC95,OT-NPA04}, but the
predictions for the production of the most exotic nuclei are still
very uncertain. The few semiempirical
models~\cite{AG-PLB74,DJM-PRC89} used to describe the data assume
Gaussian momentum distributions, characterized just by two
parameters (the mean value and the width) that may not be sufficient
to model the momentum distributions. Therefore, the measurement of
the fragment momentum distributions remains an important part of the
search for new isotopes.

\begin{figure}[b]
\includegraphics[width=0.8\columnwidth]{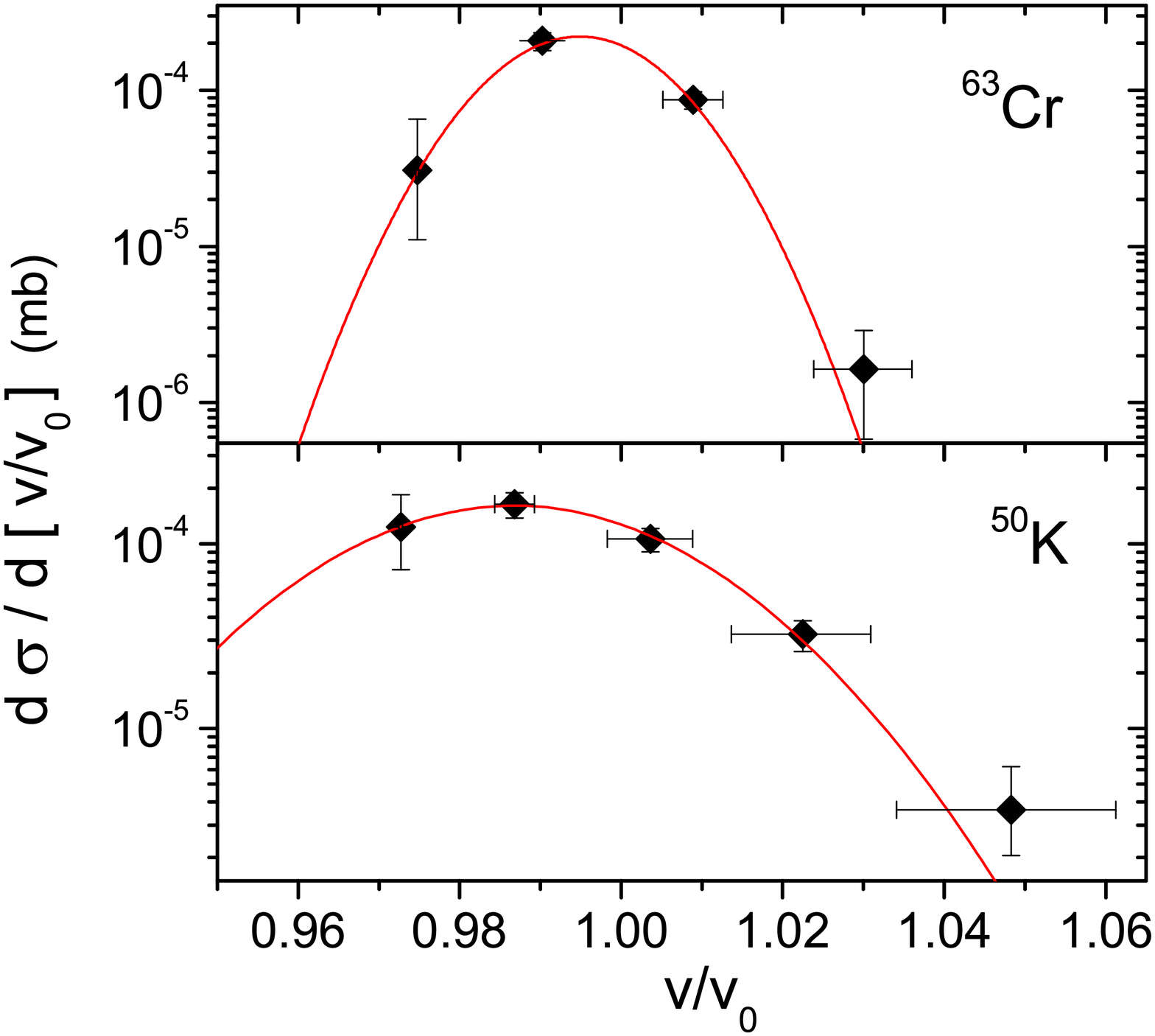}%
\caption{(Color online) Differential cross sections of $^{63}$Cr
(top) and $^{50}$K (bottom figure) fragments obtained by varying the
target thickness at one magnetic rigidity. Red solid lines represent
the fitted Gaussian functions. Horizontal errors correspond to the
velocity difference caused by production at the beginning or the end
of target.\label{Fig_Velocity}}
\end{figure}

The most straightforward and  common way to measure the momentum
distributions of projectile fragments is to scan the magnetic
rigidity of the fragment separator. A thin target is generally used
for this measurement to avoid complications from differential
energy-loss in the target (the systematic change in the kinetic
energy lost by the beam nucleus and  product nuclei in the target).
In addition, the momentum acceptance of the fragment separator is
often restricted to a small value
 for isotopes with  large yields. The acceptance is then
opened up for the smaller yields, producing fewer measurements with
larger systematic uncertainties. The total cross section is
determined by integrating these momentum distributions using some
estimate, or measurement, of the angular transmission,
e.g.~\cite{MF-PRC96}.

 In the present work a new approach to measure momentum
distributions and cross sections was used.  In contrast to the
``$B\rho$-scanning" method using one thin target, a variety of
targets with different thicknesses was used  at constant magnetic
rigidity. This method is particularly well suited to survey
neutron-rich nuclei since the less exotic nuclei are produced with
the highest yields and their momentum distributions can be measured
with the thin targets. The lighter fragments with high yields will
experience the largest differential energy losses and will fall
outside the constant momentum acceptance of the separator with the
thicker targets. Conversely, the heaviest fragments with the lowest
yields will only be produced in sufficient numbers  in the thickest
targets and will have lower differential energy losses. Extensive
simulations of the A1900/S800 tandem separator showed that this
method can provide a sensitive measurement of the mean value and
width of the momentum distribution~\cite{OT-NIMA09}.  However, as in
the $B\rho$ scanning method, the momentum distributions of some
fragments will be incomplete.

\begin{figure}
\includegraphics[width=1.0\columnwidth]{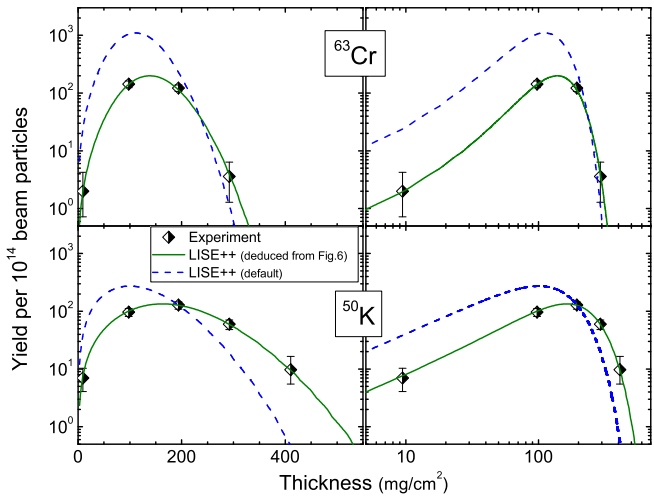}%
\caption{ (Color online) The measured yield of $^{63}$Cr (top
panels) and $^{50}$K (bottom panels) as function of target
thickness.  The lines show the \lisepp calculations with results of
fitting the velocity distributions shown in Fig.~\ref{Fig_Velocity}
(green solid line) and default model settings~\cite{DJM-PRC89} (blue
dashed line) as described in the text. Note that the left panels
have linear horizontal scales, while the right panels have
logarithmic scales to emphasize the difference between calculated
curves for the thick and thin targets. \label{Fig_Thickness}}
\end{figure}

Five targets were used to measure the momentum distributions (see
Table~\ref{Tab_runs}).  The yields of two example fragments,
$^{63}$Cr and $^{50}$K  shown in Fig.~\ref{Fig_Velocity} as a
function of reduced velocity were fitted with Gaussian functions and
the total production cross sections were inferred. The momentum
distributions for 34 isotopes were derived (indicated by the bullets
in Fig.~\ref{chart}) and integrated to deduce the cross sections. It
should be noted that at the energy of these experiments, the shape
of the fragment momentum distribution is asymmetric due to  a
low-energy exponential tail stemming from dissipative
processes~\cite{OT-NPA04}. The Gaussian function used for the
present data does not take this tail into account, but the
underestimation of the cross section is small. As one example, the
$^{58}$Ca cross section is underestimated by approximately~2\%.

A survey of all of the fitted results showed that fragments in the
heavy mass region were produced with significantly higher velocities
and slightly broader momentum distributions than the model
predictions. The model~\cite{DJM-PRC89}  used for these calculations
assumes that the energy  necessary to remove each nucleon is $E_S  =
8$~MeV, a value derived  for  fragments close to stability. Further
analysis showed that the separation energy parameter  for nuclei
observed in the present work in the region $A_P/2 \le A_F \le A_P$
exhibits a linear decrease with the number of removed nucleons:
\begin{equation} \label{Eq_Velocity}
E_S  = 8 - 11.2 \Delta A / A_P
\end{equation}
where $\Delta A = A_P - A_F$, $A_P$ is the projectile mass number,
and $A_F$ is the fragment mass number.

 The measured yields of the two example isotopes, $^{63}$Cr and $^{50}$K, are shown
as a function of target thickness in Fig.~\ref{Fig_Thickness}. The
fragment yields calculated with the \liseppsh~code~\cite{OT-NIMB08}
using parameters from Morrissey's model of the reaction with  \epax
cross sections~\cite{KS-PRC00} are shown by the dashed lines.  The
fragment yields from the same calculations with the experimental
parameters from fitting the velocity distributions (see
Fig.~\ref{Fig_Velocity}) are shown by the solid lines in
Fig.~\ref{Fig_Thickness}. Only a fair agreement is found with the
model calculations with default settings. On the other hand, the
agreement between experimental data and model calculations with
parameters produced by fitting the velocity distributions is
excellent and gives confidence that the fragment separator was set
to optimize the yield of the weakest, most neutron-rich fragments.


\subsection{Production cross section\label{secCS}}

\begin{SCfigure*}
\includegraphics[width=0.76\textwidth]{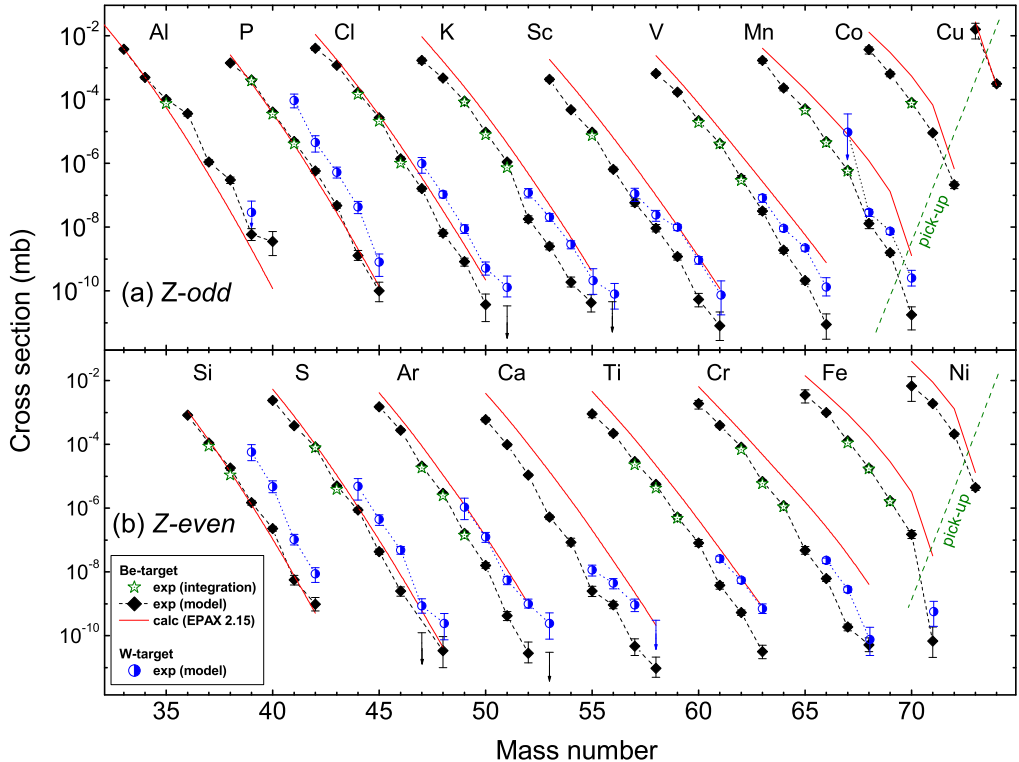}%
\caption{(Color online) Inclusive production cross sections for
fragments from the reaction of $^{76}$Ge with beryllium and tungsten
targets shown as a function of mass number. The cross sections with
the beryllium targets derived by momentum distribution integration
are shown by stars, those normalized with \liseppsh\ transmission
calculations are indicated by solid diamonds The cross sections
obtained with the tungsten target were normalized with \liseppsh\
transmission calculations. The two green dashed lines separate
nuclei that require neutron pickup in the production mechanism. The
red lines show the predictions of the \epax systematics for
beryllium (see text).
\\
\\
\label{Fig_CrossSection}}
\end{SCfigure*}

The inclusive production cross sections for the observed fragments
were calculated by correcting the measured yields for the finite
longitudinal and angular acceptances of the separator system. A
total of thirty-four cross sections with beryllium were obtained
from Gaussian functions fitted to the longitudinal momentum
distributions; these nuclei are indicated by stars in
Fig.~\ref{Fig_CrossSection}. The cross sections for all of the
remaining fragments with incomplete longitudinal momentum
distributions were obtained with estimated transmission corrections.
The angular and longitudinal transmissions were calculated for each
isotope in each setting using a model of the momentum distribution
with smoothly varying parameters extracted from the measured
parallel momentum distributions. For example, the values of the
angular, momentum and overall transmissions, as well as the wedge
selection for the central fragment in runs with a wedge are given in
Table~\ref{Tab_transmission}. The wedge selection represents the
fraction of isotopes that passed through the slits at the  focal
plane when the wedge is inserted at the intermediate dispersive
image. The  estimate of the uncertainties in the transmissions came
from the values at the  one-sigma limits of the reduced width of the
longitudinal momentum distributions scaled by the mass
loss~\cite{DJM-PRC89}:
\begin{equation} \label{Eq_Morrissey}
\sigma_{||} = \sigma_0 \sqrt{\Delta A}
\end{equation}
where $\Delta A$ was defined above, and $\sigma_0=105\pm15$~MeV/c.
Details of transmission  calculations, the analysis of their
uncertainties, as well as the general analysis of momentum
distributions taken from various target thicknesses will be
described in a subsequent paper~\cite{OT-NIMA09}.

\begin{table}[b]

\caption{ Transmissions during runs with a
wedge}\label{Tab_transmission}
\begin{tabular}{|c|c|ccc|c|}
\hline
  ~Data~   & {\small{Fragment}}  &     \multicolumn{ 4}{|c|}{Transmission(\%)} \\
\cline{3-6}
      set  &  {\small{of}}       &     Momentum  &    Angular &  Wedge     &    ~Total~  \\

           &  {\small{interest}} &               &            & selection  &\\
\hline
         6 &        $^{43}$S  &    $11\pm3$ &    $70^{+7}_{-8}$&     $75^{+10}_{-14}$&     $6\pm2$ \\
         7 &        $^{43}$S  &    $16\pm3$ &    $69^{+7}_{-8}$&     $74^{+10}_{-14}$&     $8\pm2$ \\
         8 &        $^{66}$V  &    $70\pm3$ &    $99^{+1}_{-2}$&     $83^{+8}_{-13}$ &    $58^{+6}_{-10}$\\
         9 &       $^{66}$V   &    $69\pm3$ &    $99^{+1}_{-2}$&     $83^{+9}_{-13}$&    $56^{+6}_{-9}$ \\
        10 &       $^{54}$Ar  &    $47\pm6$ &    $88^{+4}_{-6}$&     $79^{+9}_{-14}$&    $33^{+6}_{-8}$ \\
        11 &       $^{54}$Ar  &    $30\pm3$ &    $86^{+7}_{-6}$&     $78^{+10}_{-14}$&   $20^{+3}_{-4}$  \\
        12 &       $^{59}$Ca  &    $32\pm3$ &    $93^{+3}_{-5}$&     $80^{+9}_{-14}$&   $24^{+3}_{-5}$ \\
        13 &       $^{58}$Ca  &   $41^{+2}_{-1}$ &    $93^{+3}_{-5}$&     $80^{+9}_{-14}$ &   $31^{+4}_{-6}$\\
\hline
\end{tabular}
\end{table}

The cross sections obtained for all of the fragments observed in
this experiment are shown in Fig.~\ref{Fig_CrossSection} along with
the predictions of the \epax parameterization. The systematic errors
associated with the target thickness, beam current measurements, and
the widths of the momentum slits have been included in the error
bars and were 11\% for the first data set, and approximately 2\% for
production data sets~(8-13). For those isotopes that relied on
transmission calculations, the weighted mean of all measured yields
was used to obtain the ``model-based'' cross section (shown by solid
diamonds in Fig.~\ref{Fig_CrossSection}). The uncertainties in these
cases included the statistical, the systematic and the transmission
uncertainties. As can be seen in Fig.~\ref{Fig_CrossSection}, the
model-based cross sections are in good agreement with those produced
by integrating the measured longitudinal momentum distributions as
was shown above for the examples $^{63}$Cr and $^{50}$K.


\begin{SCfigure*}
\includegraphics[width=0.8\textwidth]{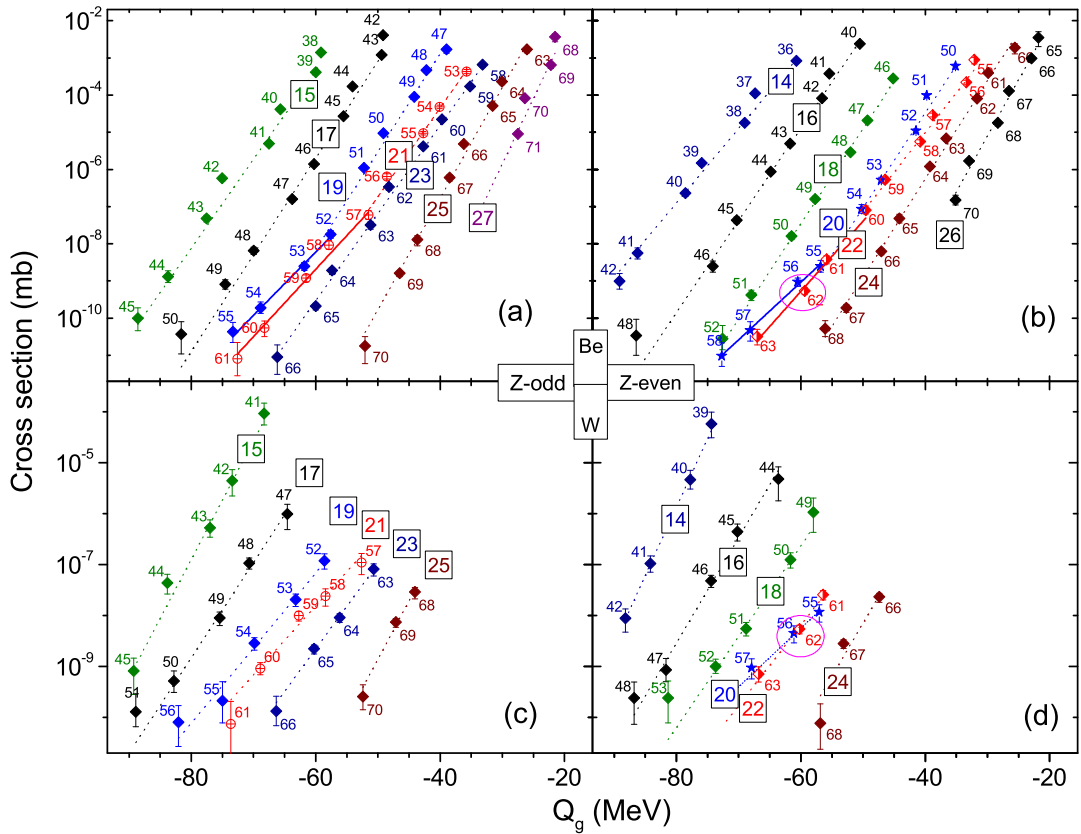}%
\caption{(Color online) Cross sections for the production of
neutron-rich nuclei with (a,c) odd atomic numbers and (b,d) even
atomic numbers, with (a,b) a beryllium target~\cite{OT-PRL09} and
(c,d) tungsten target, respectively. See text for explanation of
$Q_g$ and the lines. The cross sections for $^{62}$Ti at the center
of the proposed new island of
inversion~\protect{\cite{BAB-PPNP01}} are circled. \\
\label{Fig_Qg}}
\end{SCfigure*}

The runs with the tungsten target employed  stripper foils (Be and
C) placed directly behind the target which should be taken into
account for the determination of the cross sections. Such a
correction can easily be made for the most neutron-rich isotopes
(for example $^{59-61}$Sc) for which the ratio of the observed cross
sections $R=\sigma_W / \sigma_{Be}$ exceeds $\approx$10. Cross
sections with the tungsten target for isotopes with $R\approx$1 can
be reliably obtained if the corresponding cross section with the
beryllium target was measured with a small overall uncertainty (for
example $^{57,58}$Sc).  On the other hand, the division of the
yields of $^{54-56}$Sc measured with the tungsten target and a
stripper between the two target materials was very uncertain. In
addition, reliable lower limits could not be established for the two
interesting isotopes $^{40}$Al and $^{67}$Mn due to large
uncertainties in the transmission. The cross sections for the
production of some other isotopes (e.g., $^{51}$Cl and $^{56}$K)
with the beryllium targets, needed to correct the measured cross
sections with the tungsten target, had to be extrapolated. Finally,
the cross sections needed for the correction of the data with the
carbon stripper (9$^{th}$ data set in Table~\ref{Tab_runs}) were
calculated by normalizing the beryllium cross sections by the ratio
of the total reaction cross sections $\sigma_{C}/ \sigma_{Be}=1.05$
from the \epax parametrization.

The compilation of results in Fig.~\ref{Fig_CrossSection} clearly
shows a larger (sometimes exceeding a factor of ten) cross section
with the tungsten target for the production of very neutron-rich
isotopes of the elements \protect{$16\le Z\le 25$} at this
projectile energy. The geometrical factor (based on \epaxsh) is only
1.9. Similarly large ratios were  recently obtained with a $^{48}$Ca
primary beam at a very similar beam energy~\cite{OT-PRC07}. Models
of nuclear reactions used for counting rate estimates, like the
intranuclear-cascade plus evaporation model \cite{DJM-PRL79} or
abrasion-ablation in \liseppsh~\cite{OT-NIMB03} do not reproduce the
low yields of exotic nuclei observed in this work. The predictions
of the  \epax parameterization for reactions with beryllium, shown
by the solid lines in Fig.~\ref{Fig_CrossSection}, reproduces the
measured cross sections rather well in the region  \protect{$13\le
Z\le 16$}. The data for the elements with $Z > 16$ generally fall
below the predicted values but have a similar slope.


\subsection{Q$_g$ systematics\label{secQg}}

\begin{figure}[b]
\includegraphics[width=0.95\columnwidth]{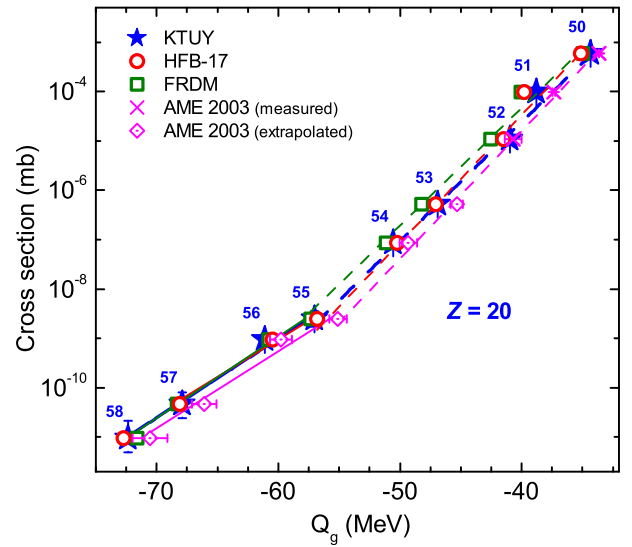}%
\caption{ (Color online) Production cross sections  of neutron-rich
calcium isotopes with  a beryllium target as a function of $Q_g$
calculated with different mass models, see text.}
\label{Fig_MassModels}
\end{figure}

The production cross sections for the most neutron-rich projectile
fragments have been  previously shown to have an exponential
dependance on $Q_g$ (the difference in mass-excess of the beam
particle and the observed fragment)~\cite{OT-PRC07,OT-PRL09}.
However, it is important to note that the actual masses of the very
neutron-rich nuclei, needed for this calculation, have not been
measured and are only available from models.  There are many
differences among mass models therefore deviations from the
predicted yield using a given mass model might be able to identify
missing features in that mass model. For example, significant
deviations of the real masses can be expected from the model
predictions if a new island of inversion near $^{62}$Ti
\cite{BAB-PPNP01} (similar to the island of inversion observed near
$^{31}$Na) is present but is omitted from the models. In a
shell-model picture, the ground states of nuclei in the new island
of inversion would be dominated by intruder configurations
corresponding to neutron particle-hole excitations across the $N=40$
sub-shell gap into the $g_{9/2}$~\cite{BAB-PPNP01}. As a test of
this suggestion, the cross sections for the production of all nuclei
observed with the beryllium and tungsten targets are shown in
Fig.~\ref{Fig_Qg} where the abscissa, $Q_{\text{g}}$, is the
difference between the mass of the ground state of the projectile
and the observed fragment. In this case the masses were taken from
Ref.~\cite{KTUY-PTP05}. This empirical mass model (KTUY) provides an
excellent systematization of the variation of the data, with the
logarithm of the cross sections for each isotopic chain falling on
an approximately straight line.

The cross sections for each isotopic chain were fitted with the simple expression:
\begin{equation}\label{Eq_Qg}
 \sigma(Z,A) = f(Z)\exp{(Q_{\text{g}}/T)},
\end{equation}
where  $T$ is an effective temperature extracted from the slope.
Nearly all the data from the beryllium targets can be fitted with
the single value of $T= 1.8$~MeV, but slightly better fits are
obtained by finding the optimum $T$ for each element. However, the
heaviest isotopes of elements in the middle of the distribution
($Z=$~19, 20, 21, and 22) appear to break away from the
straight-line behavior.  These heaviest four or five isotopes of
these elements were found to have a shallower slope (shown by the
solid lines) or enhanced cross sections relative to the model
prediction.

The systematic variation of the production cross sections as a
function of $Q_g$ was checked with several other well-know mass
models and essentially the same behavior was observed. Note that the
differences among the mass models are small compared to the abscissa
in Fig.~\ref{Fig_Qg}. Specifically, models based on the
Hartree-Fock-Bogoliubov method HFB-8, HFB-9~\cite{MS-NPA03}
including the very recent HFB-17 version~\cite{SG-PRL09}, the finite
range droplet model (FRDM)~\cite{PM-ADNDT95}, and the 2003 Atomic
Mass Evaluation (AME2003)~\cite{GA-NPA03} with shell crossing
corrections that were developed in the \lisepp
framework~\cite{OT-PRE02} were found to give similar results. The
different behavior of the calcium isotopes with the beryllium target
can be seen more clearly in Fig.~\ref{Fig_MassModels}. The same
effect is present in the same mass region with the tungsten target.

\begin{figure}
\includegraphics[width=1.0\columnwidth]{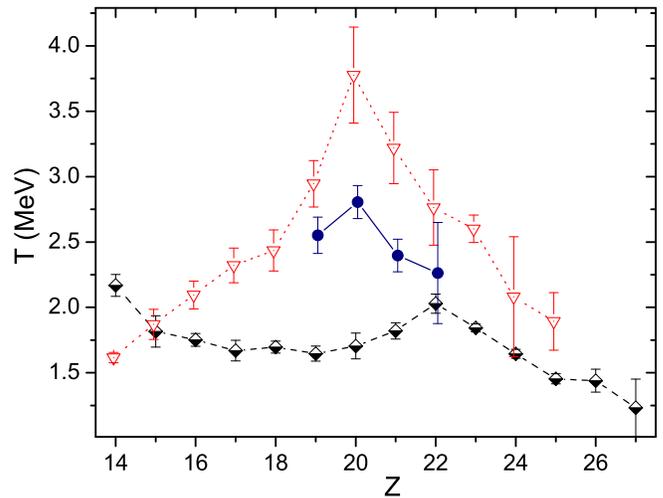}%
\caption{(Color online) Values of the inverse slope parameter, $T$,
from the best fit of Eq.~\ref{Eq_Qg} to the experimental cross
sections in Fig.~\ref{Fig_Qg} shown as a function of atomic number.
Open triangles connected by dotted lines are from the tungsten
target, half-filled diamonds connected by dashed lines are from the
beryllium target and the filled circles connected by solid lines are
from the heaviest isotopes from the beryllium target in
Fig.~\ref{Fig_Qg}.} \label{Fig_Temperature}
\end{figure}

The variation of the individual fitted values of the inverse slope
parameter, $T$, for products from both targets is shown as a
function of atomic number in Fig.\ref{Fig_Temperature}. The inverse
slopes of the cross sections from the beryllium target connected by
dashed (solid) lines in Fig.~\ref{Fig_Qg} are indicated by the
half-filled black diamonds and dashed lines (the filled circles and
solid lines) in Fig.~\ref{Fig_Temperature}. Similar for the tungsten
target the inverse slopes of the measured cross sections are shown
by the open red triangles connected by dotted lines in
Fig.~\ref{Fig_Temperature}, but the range of cross sections was more
limited and only one slope parameter was fitted to their data.
Fig.~\ref{Fig_Qg} clearly shows that there is a general increase in
 $T$ for all of the heavy isotopes
observed with $Z=$~19, 20, 21, and 22.

One possible explanation for the shallower slope of the heaviest
nuclei of these isotopes is that these nuclei are more bound (i.e.,
less negative $Q_g$) than predicted by the mass models. A source of
stronger binding can be deformation. In a shell-model framework, the
wave functions of the ground and low-lying excited states of nuclei
in the new island of inversion around $^{62}$Ti would be dominated
by neutron particle-hole intruder excitations across the $N=40$
sub-shell gap, leading to deformation and shape coexistence. Thus,
mass measurements of these nuclei, although difficult, would be
especially interesting.


\section{Summary\label{Summary}}

The present study of the fragmentation of a $^{76}$Ge beam at
132~MeV/u provided evidence for the production of fifteen previously
unobserved neutron-rich isotopes~\cite{OT-PRL09}. The momentum
distributions and cross sections for a large number of neutron-rich
nuclei produced by the $^{76}$Ge beam were measured by varying the
target thickness in a two-stage fragment separator with a fixed
momentum acceptance. The longitudinal momentum distributions of 34
neutron-rich isotopes of the elements with \protect{$13\le Z\le 27$}
were compared to models that describe the shape and centroid of
fragment momentum distributions.  New parameters for the
semiempirical momentum distribution model~\cite{DJM-PRC89} based on
the measured momenta were obtained. The production cross sections
support the previous analysis on the basis of the $Q_g$ function but
the most neutron-rich nuclei of elements with $Z=19$ to 22 are
produced with an enhanced rate compared to the systematics. This
trend was previously reported for fragmentation on a beryllium
target~\cite{OT-PRL09} and was found to also be present with the
tungsten target.  The enhanced yields lie in the region that was
previously predicted to be a new island of inversion in which an
intruder shell-model state becomes the ground state.

\begin{acknowledgments}
The authors would like to acknowledge the operations staff of the
NSCL for developing the intense $^{76}$Ge beam necessary for this
study. This work was supported by the U.S.~National Science
Foundation under grant PHY-06-06007.
\end{acknowledgments}


\eject
\appendix
\section{Identification of heavy ions with energy losses\label{secPID}}

The magnetic rigidity ($B\rho$) of each particle was obtained by
combining an angle measurement provided by two PPAC's in the
dispersive plane of the S800 analysis beam line in combination with
NMR measurements of the dipole fields: $B\rho$ = $B\rho_0  (1 +
\delta)$, where  $B\rho_0$ is the magnetic rigidity of the dipole at
the central axis, and $\delta= \Delta p / p$ is the fractional
deviation of the particle from the central rigidity (for a constant
charge $q$). The mass-to-charge ratio $A/q$ is given by the
expression:

\begin{equation}
\label{Eq_aq} A/q =  \frac{B\rho}{ \beta \gamma} \frac{e}{ u c }
\end{equation}

where $e$ is the elementary charge, $u$ is the atomic mass unit,
$\beta$ is the velocity of the ion relative to the speed of light
$c$, and $\gamma$ is the Lorentz factor. The velocity of the ions
was determined from the time-of-flight (TOF). However, for the TOF
measurement between the A1900 focal-plane scintillator and the
second PIN detector (see Table~\ref{Tab_tof} and
Fig.~\ref{Fig_setup}), there were several materials in the path of
the beam leading to different  ion velocities along the beam line.
In order to calculate the $A/q$ ratio in this case with
Eq.\ref{Eq_aq}, it was necessary to introduce a reduced $B\rho_a $
value that corresponds to the average velocity $\beta_{aver}= L /
(cT)$, where $T$ is the TOF between two timing detectors. The
time-of-flight value $T$ can be written as $ T = \sum_i^N L_i /
(c\beta_i) $, where $N$ is the number of stages where the ion
velocity changed due to the passage through material, and $L_i$ is
the corresponding flight path at a given velocity.

At intermediate beam energies when the ion's velocity is not
significantly changed along the path (e.g. only thin materials are
inserted in the beam)  one can use the reduced magnetic rigidity as:
\begin{equation} \label{Eq_brhoreduced}
B\rho_a =   k  L / \sum_i^N(L_i/B\rho_{0i})
 \end{equation}
 where $k$ is the momentum deviation measured before the
last flight stage, and $B\rho_{0i}$ is the magnetic rigidity
corresponding to the $i^{th}$-stage  on the optical axis. At these
energies ($\beta=0.4$) it is possible to omit the $\gamma$-factor in
Eq.~\ref{Eq_aq} because  the ratio of $(\delta\beta/\beta) /
(\delta\gamma/\gamma)=5.2$, whereas it is equal to 1.0 at
$\beta=0.71$.

Calculations using Eq.~\ref{Eq_brhoreduced} for $^{58}$Ca fragments
(Data set~13, see Table~\ref{Tab_runs}) with the TOF measured
between the FP scintillator and the  2$^{nd}$ PIN detector (see
Table~\ref{Tab_tof}) were compared to \lisepp Monte Carlo
simulations. A maximum difference of $d\beta = 0.06\%$ occurs for
ions farthest from the central rigidity ($\Delta p/p=2.5\%$). This
results in a mass resolution for $^{58}$Ca  of about 0.04 units.

The charge state ($q$) of the ion was evaluated with a relation
based on the total kinetic energy (TKE), velocity,  and magnetic
rigidity:
\begin{equation} \label{Eq_Q}
q = \frac{TKE}{u(\gamma - 1)(A/q)}
\end{equation}
where TKE is calculated as a sum of the energy loss values in each
of the Si detectors. The velocity before the Si telescope was
calculated from the reduced magnetic rigidity and the central
magnetic rigidity of the last dipole in the beam line
($B\rho_{0f}$):

\begin{equation} \label{Eq_betaf}
\beta = \frac{k ~ \beta_{aver} ~ B\rho_{0f}}{B\rho_a}
\end{equation}

The atomic number was determined from the combination of energy loss
($\Delta E$) and velocity values according to the Bethe formula:
\begin{equation} \label{Eq_Z}
 Z \approx \sqrt{ \frac{\Delta E}{\left [ MeV \right ] } \left (  \frac{1}{\beta^2}
 \log{\frac{5930}{1/\beta^2-1}}-1 \right )}
\end{equation}

\end{document}